\documentclass{elsart}

\usepackage{graphicx}

\begin{document}

\begin{frontmatter}

\title{Optimal adaptive performance and delocalization in
  NK fitness landscapes}

\author[SaoCarlos]{Paulo R. A. Campos},
\author[JPL,Caltech]{Christoph Adami}, and
\author[Caltech]
{Claus O. Wilke}

\address[SaoCarlos]{Instituto de F\'{\i}sica de S\~ao Carlos, Universidade de
  S\~ao Paulo, Caixa Postal 369, 13560-970 S\~ao Carlos SP, Brazil}

\address[JPL]{Jet Propulsion Lab, MS 126-347, California Institute of
  Technology, Pasadena, CA 91109}

\address[Caltech]{Digital Life Lab, Mail Code 136-93
California Institute of Technology, Pasadena, CA 91125}



\begin{abstract}
We investigate the evolutionary dynamics of a finite population of
sequences adapting to NK fitness landscapes. We find that, unlike in
the case of an infinite population, the average fitness in a finite
population is maximized at a small but finite, rather than vanishing,
mutation rate. The highest local maxima in the landscape are visited
for even larger mutation rates, close to a transition point at
which the population {\it delocalizes} (i.e., leaves the fitness peak at
which it was localized) and starts traversing the sequence
space. If the mutation rate is increased even further, the population
undergoes a second transition and loses all sensitivity to fitness
peaks. This second transition corresponds to the standard error
threshold transition first described by Eigen.  We discuss the
implications of our results for biological evolution and for
evolutionary optimization techniques.
\end{abstract}

\end{frontmatter}

\section{Introduction}\label{sec:level1}

The evolution of a finite population in a rugged, multi-peaked fitness
landscape is an important area of research in theoretical
biology. Although thoroughly studied over the last 15 years, we are
still far from a complete understanding of the intricate dynamics that
unfold. Most analytical results deal with very simple fitness
landscapes, such as the flat landscape~\cite{DerridaPeliti91}, the
single peak
landscape~\cite{NowakSchuster89,CamposFontanari98,CamposFontanari99},
multiplicative~\cite{HiggsWoodcock95,WoodcockHiggs96} or
additive~\cite{PruegelBennett97} landscapes, or other landscapes that
contain a high degree of symmetries, such as the Royal Road
landscape~\cite{vanNimwegenetal99a}. Numerically, one may study more
complicated situations, such as RNA folding
landscapes~\cite{FontanaSchuster98,Huynenetal96} or self-replicating
computer programs~\cite{Adami98}. Here, we are interested in the
family of NK landscapes~\cite{KauffmanLevin87,Kauffman92}. The NK
landscapes are interesting because their ruggedness can be tuned, from
a single smooth peak to a completely random landscape, so that the
influence of ruggedness on the dynamics of an evolving population can
be studied systematically.

Traditionally, NK landscapes have been studied in the context of an
adaptive walker, which is essentially a population of size one. If the product
of population size $M$ and mutation rate $u$ is small, $uM \ll 1$, the
adaptive walk is a reasonable approximation of the full population dynamics.
For larger products $uM$, however, quasispecies effects such as error
thresholds~\cite{Eigenetal89} or selection for mutational
robustness~\cite{vanNimwegenetal99b,Wilke2001} are to be expected. Here,
we are mainly interested in the change of the population dynamics as the
mutation rate is increased, so that $uM$ cannot be considered small.

We investigate the adaptive performance of a finite population on an NK
landscape as a function of the mutation rate. Depending on whether we
consider the mean population fitness or the maximum fitness in the
population, we find different optimal mutation rates. The mean fitness
is optimized at a mutation rate that is just sufficiently high to
prevent a complete collapse of the population into isogeny (a single
genotype). At such a mutation rate, sequence diversity is increased to
a value which allows a population to explore the genotype space more
efficiently, without losing too much fitness via accumulating
deleterious mutations. The maximum fitness, on the other hand, is
highest when the mutation rate is so high that the population is on
the verge of delocalization. At such a mutation rate, a
population is just barely able to maintain the information discovered
about the landscape, while at the same time it can explore genotype
space at an optimal speed.

\section{The NK model}
The NK model was introduced by Kauffman~\cite{KauffmanLevin87} in order to
study the influence of landscape ruggedness on adaptive evolution. The model
is similar to certain physical models of spin-glasses, in particular to the
random-energy model~\cite{Derrida81}.

The model is defined as follows. We assume that each organism in the
population is composed of $N$ segments, or genes, each of which can be
in one of two possible states, designated by 0 or 1. The fitness value
of an organisms is then given by the average value of the selective
contribution of each of its $N$ genes,
\begin{equation}
F = \frac{1}{N}\sum_{i=1}^{N} f_{i},
\end{equation}
where $f_{i}$ denotes the contribution of gene $i$ to the fitness
value and is a function of its own state and of the state of $K$ other
genes randomly chosen from the remaining $N-1$ ones. The values $f_i$
itself are drawn randomly from a uniform distribution on the interval
$(0,1]$. All random variables in the NK model are quenched, i.e., they
are chosen once at the beginning of a simulation run, and then held
fixed. Since the functions $f_i$ depend on the state of $K$ different
genes at a time, the genes in the NK model are not independent, they
interact (i.e., the model assumes \emph{epistasis} between genes). By
changing the number of genes $K$ participating in the epistatic
interaction, we can shape the fitness landscape. The 
parameter $K$ controls the ruggedness of the landscape. For
small values of $K$ the landscape is smooth, becoming increasingly
rugged for higher values of $K$.

\begin{figure}
\centerline{
\includegraphics[width=9cm]{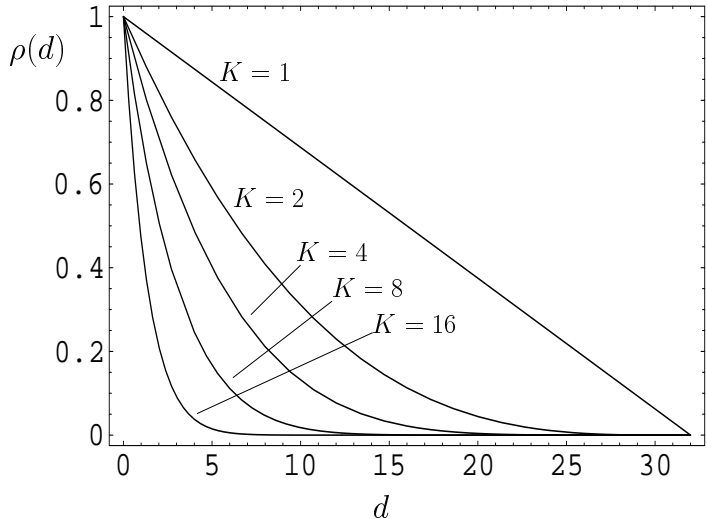}
}
\caption{Auto-correlation function in NK landscape with $N=32$. 
\label{fig:corrfunc}}
\end{figure}

The influence of the parameter $K$ on the ruggedness of the NK landscape can
be visualized with the aid of the auto-correlation function, derived
in the Appendix:
\begin{equation}
  \rho(d) = \frac{(N-K)!(N-d)!}{N!(N-K-d)!}\,.
\end{equation}
A fast decay of $\rho(d)$ indicates a high degree of ruggedness, because in
that case even sequences that are only a few mutations apart contain hardly
any information about each other. If $\rho(d)$ decays slowly, on the other
hand, information is preserved over large distances in genotype space, which
is only possible if the fitness peaks in the landscape are very broad and
smooth.  In Fig.~\ref{fig:corrfunc}, we have displayed $\rho(d)$ for $N=32$
and various choices of $K$. We find that for $K=1$, the auto-correlation
function decays very slowly, in agreement with a smooth fitness landscape. As
$K$ is increased, $\rho(d)$ decays increasingly faster, and the landscape
becomes more rugged. In the extreme case of $K=N-1$ (not displayed), $\rho(d)$
decays to zero at $d=1$, indicating that the landscape has become completely
random at that point.

In order to simulate a population evolving in an NK landscape, we use the
following algorithm, which has been used previously by Sibani and
Pedersen~\cite{SibaniPedersen99}, and is described in detail in
Ref.~\cite{Pedersen99}. First, we select 50\% of the $M$ sequences in the
population for reproduction. An organism is selected with probability
proportional to its fitness, so that the most fit individuals contribute most
to the composition of the population in the next time step. After
reproduction, the population has then reached a size of $1.5 M$. Now, we
randomly remove one third of the individuals, so that the population after
replication and removal consists again of $M$ individuals. The replication
mechanism is assumed to be imperfect, and the probability of mutation for a
single gene is given by the rate $u$.  Recombination is not taken into
account in the present work.

\section{Results}

Our main interest in this work lies in identifying the mutation rate
at which a finite population performs ``best'' in an NK landscape. By
``best'', we mean that in equilibrium, and averaged over many
independent runs, either the mean or the maximum fitness of the
population is maximized. For an infinite population, this question is
trivial, since the mean fitness in equilibrium is always maximized at
zero mutation rate, and can only decay as the mutation rate
increases. For a finite population, however, a mutation rate that is
too small leads to a premature standstill in the progress of
adaptation, as the population gets trapped in local optima. A mutation
rate that is too large, on the other hand, drives the population away
from very narrow but high peaks. The optimum mutation rate therefore
strikes the right balance between the potential for barrier crossing and
the risk of destabilizing fluctuations once a peak has been reached.

\begin{figure}[bp]
\vspace{.3in}
\centerline{
\includegraphics[width=8cm,angle=270]{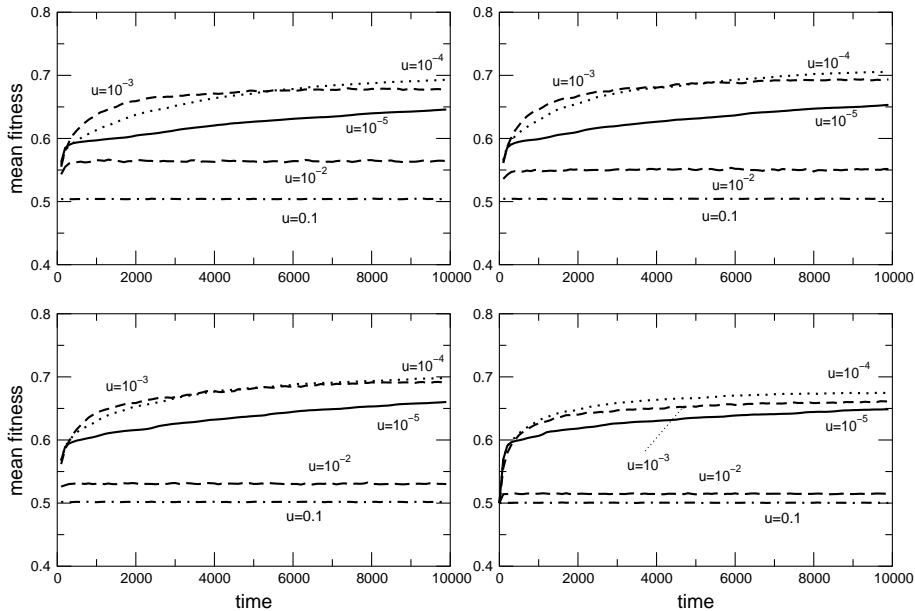}
}
\caption{Mean fitness as a function of time, averaged over 50 independent
  simulation runs, for $N=64$ and $M=5000$. From left to right, and top to
  bottom, $K=2$, $K=4$, $K=8$, and $K=16$.}
\label{fig:kauff-2}
\end{figure}

Figure \ref{fig:kauff-2} shows the mean fitness as a function of time,
for various values of $K$ and the mutation rate $u$. The results for
different values of $K$ are very similar. We observe an optimal
mutation rate (in terms of mean fitness) around $u=10^{-4}$ (a genomic
mutation rate of $\mu=Nu=0.0064$). For mutation rates below that
value, the mean fitness grows much more slowly.  If we increase the
mutation rate beyond this value, we find that while the initial
adaptation during the first 1,000-2,000 time steps is significantly
faster, mean fitness actually drops.  This is due to the more
efficient exploration of genetic space that a higher mutation rate
entails.  Initially, when the population starts out in a valley of low
fitness, the higher genetic diversity of a population at a higher
mutation rate results in more frequent discoveries of higher fitness
genotypes. However, when equilibration is reached, the high mutation
rate creates a constant influx of deleterious mutations, which reduce
the mean fitness in equilibrium.

In order to obtain a more detailed picture of this dynamics, we
studied the case of $K=8$ more thoroughly. We recorded mean and
maximum fitness after $20,000$ time steps, averaged over 50
independent runs. In addition, we recorded the average Hamming
distance to the population's consensus sequence, in order to obtain a
deeper understanding of the population structure that forms at
different mutation rates.

\begin{figure}
\centerline{
\includegraphics[width=9cm]{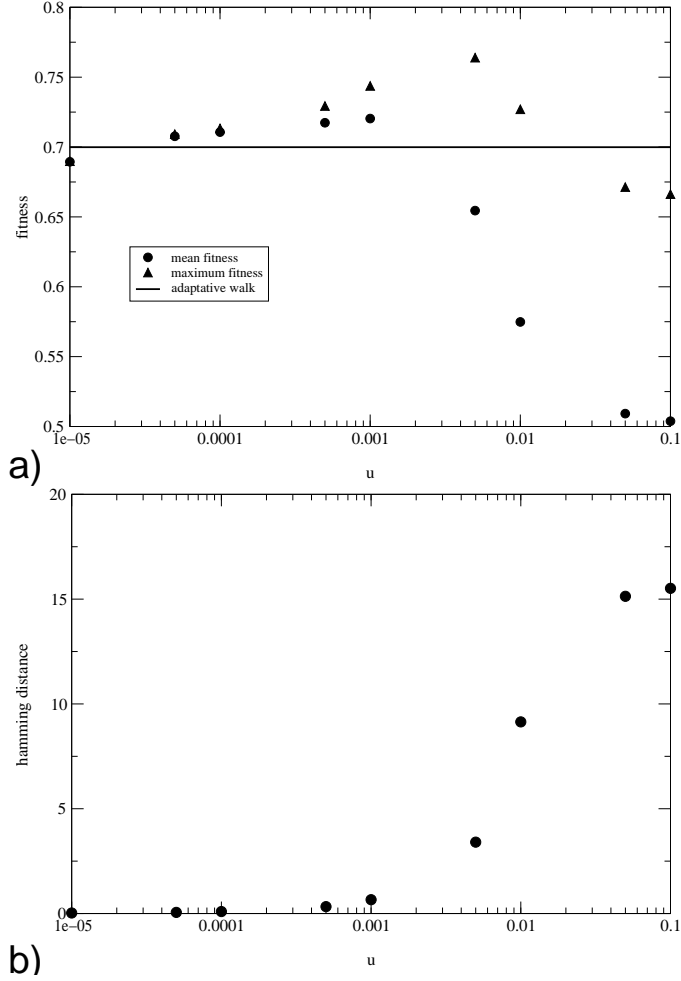}
}
\caption{(a) Maximum and mean fitness after 20,000 time steps, averaged over 50
  independent runs, as a function of the mutation rate, for $N=32$, $K=8$, and
  $M=1,000$. (b) Hamming distance
  to consensus sequence in the same simulations.}
\label{fig:meanmax-and-hamming}
\end{figure}

Figure \ref{fig:meanmax-and-hamming}a) shows mean and maximum fitness in the
population as a function of the mutation rate $u$. For comparison, we have
also displayed the average height of local maxima in the landscape, obtained
as the average over the final fitness of repeated adaptive walks starting from
random positions in the landscape. The mean fitness reaches its maximum value
around $u=0.001$ ($\mu=0.032$), and the maximum fitness around $u=0.005$ ($\mu=0.16$). At $u=0.1$, the
mean fitness has reached the value 0.5, which corresponds to the average
fitness of an arbitrary sequence in a NK landscape. At this mutation rate,
selection has ceased to have any influence on the population.

For a very low mutation rate, $u=10^{-5}$, we observe that both the
mean and the maximum fitness lie below the average height of local
maxima in the landscape. However, we expect them to lie exactly on the
average height of local maxima in perfect equilibration, because a
finite population behaves like an adaptive walker if the mutation rate
is sufficiently low. The discrepancy we observe is caused by the
finite time of the experiment (20,000 time steps). The lower the
mutation rate, the longer it takes until a local maximum is reached,
because the rate at which advantageous mutations are discovered is
directly proportional to $u$ for small $u$. In the particular case of
$u=10^{-5}$, the time allotted for the simulations was not sufficient
to allow complete equilibration.

For mutation rates above $u=10^{-4}$, both mean and maximum fitness lie above
the height of the local maxima. This demonstrates that a finite population,
evolving at the appropriate mutation rate, can perform truly better than an
adaptive walk, due to the fact that it can cross fitness barriers in
situations where an adaptive walker would simply get stuck in a local
sub-optimum.

Figure \ref{fig:meanmax-and-hamming}b) shows the mean Hamming distance
to the consensus sequence as a function of the mutation rate. For
small $u$, we witness the collapse of the population to an isogenic
one, reflected in an average Hamming distance of zero. In that regime,
the dynamics of the population is only determined by the rate at which
advantageous mutations are found, as discussed above. For larger $u$,
the Hamming distance between sequences increases. At about $u=0.005$,
which is the value at which a population discovers the highest local
optima, the average Hamming distance to the consensus sequence is
already about three. Beyond $u=0.005$, the Hamming distance increases
quickly, until it reaches approximately 16 at $u=0.1$. The value of 16
is exactly half the sequence length $N=32$, which means that the
population is completely random at this point. This is in agreement
with the mean fitness of 0.5 that we find at this mutation rate. At
$u=0.1$, the population transitions from order to disorder.  This
transition corresponds to the error threshold, which was first
observed by Eigen~\cite{Eigen71}, and later found in a large variety
of different evolutionary
settings~\cite{SwetinaSchuster82,Tarazona92,Hermissonetal2001,FranzPeliti1997,AlvesFontanari1997}.

\begin{figure}
\centerline{
\includegraphics[width=8cm,angle=270]{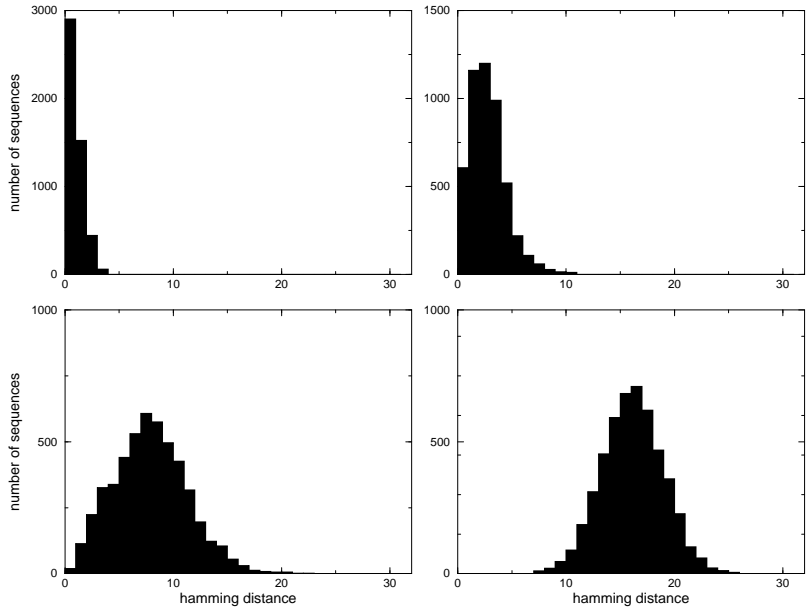}
}
\caption{Histogram for the Hamming distance between the sequences in
the population and the consensus sequence at time $t=20,000$. In these
simulations $M=5000$, $N=32$ and $K=8$. The probability of mutation
per digit is: (a) $u=0.001$, (b) $u=0.005$, (c) $u=0.01$ e (d)
$u=0.1$.}
\label{fig:histo-1}
\end{figure}

\begin{figure}
\centerline{
\includegraphics[width=8cm,angle=270]{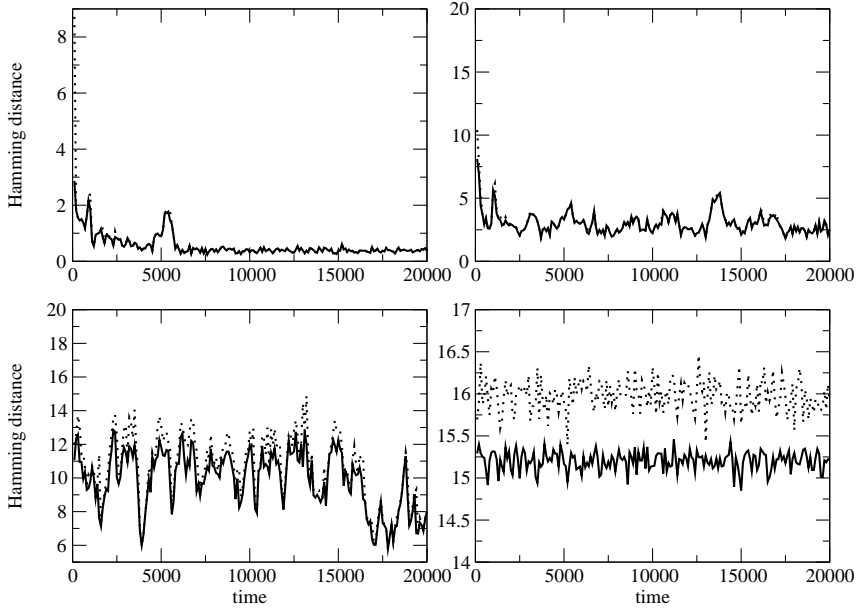}
}
\caption{The probability of mutation per digit is: (a) $u=0.001$, (b)
$u=0.005$, (c) $u=0.01$ e (d) $u=0.1$.}
\label{fig:consensus}
\end{figure}

Figure~\ref{fig:histo-1} shows the population structure at various
mutation rates in more detail, as a histogram of the sequences'
distances to the consensus sequence. We find that for $u=0.005$, the
rate at which the population discovers the highest local optima, the
consensus sequence still makes up a significant proportion of the
population, although the population's center has already moved away
from the consensus sequence, to a distance of about two. For a
slightly higher mutation rate, $u=0.01$, the population's center moves
even further away from the consensus sequence, which in turn is
represented by only 1 or 2 \% of the sequences in the population or
else disappears completely. A more detailed analysis of
this regime reveals that in addition to being almost extinct, the
consensus sequence also starts to wander about in sequence space. This
can be seen in Fig.~\ref{fig:consensus}, where we have displayed the
average distance to the consensus sequence and the average distance to
the consensus sequence of 100 time steps past, as a function of
time. While for $u=0.005$ and lower, the two curves lie exactly on top
of each other, there are significant deviations between the two curves
for larger $u$, which shows that the consensus sequence moves for
these mutation rates. However, the population has not yet crossed the
error threshold at $u=0.01$, since its structure is still clearly
different from complete randomness, and the mean fitness is still
above 0.5. Instead it flees the fitness peak it used to inhabit
(delocalizes) and settles on many adjacent, most likely
``flatter''~\cite{Wilkeetal2001} ones. The population thus undergoes
an additional transition prior to experiencing the error
threshold. This transition is a purely stochastic effect (i.e., it
depends crucially on a finite population size). We will refer to it as
the delocalization transition, since the population becomes
delocalized and starts to drift through sequence space,
re-localizing on adjacent (lower fitness) maxima. A similar transition
was previously observed by Bonhoeffer and
Stadler~\cite{BonhoefferStadler93} in two different landscapes, the
Sherrington Kirkpatrick spin glass and the Graph Bi-partitioning
landscape, and seems to be a generic finite population effect on
rugged fitness landscapes.

\section{Discussion}

In the present work, we have considered both mean and maximum fitness
as criteria for optimal adaptive performance. Which one of the two is
the more appropriate depends on the context. In biological evolution,
the mean population fitness is what determines long-term evolutionary
success. A population centered around a particularly high local
maximum can easily be driven to extinction if it carries a high load
of deleterious mutations and has to compete with another population
that has a higher mean, but lower maximum
fitness~\cite{Wilkeetal2001}. In evolutionary optimization, on the
other hand, we are interested in a single particularly good solution,
and the mean fitness is rather meaningless. In that context,
optimization of the maximum fitness is more interesting. In the
following, we will discuss our results both in the context of
biological evolution and evolutionary optimization. We begin with
biological evolution.

It has been conjectured by Eigen~\cite{Eigen86} that optimal
adaptation is realized at the verge of the error threshold, and that
natural populations should therefore evolve towards the error
threshold. Here, we have found that this is clearly not the case in NK
landscapes. Rather, the mean fitness (which is the relevant quantity
for natural populations, see above) is maximized at mutation rates
that generate only a moderate sequence diversity, far away from the
error threshold. Maximum rate of evolution also occurs before the
error threshold, {\it and} before the loss of the consensus
sequence. The true error threshold, when sequences diffuse through
sequence space without relocalization, occurs at much higher mutation
rates, after the consensus sequence is long gone.

The delocalization phase for mutation rates below the error transition
corresponds to the situation of Muller's ratchet in classical
population genetics~\cite{Muller64,Felsenstein74}. Muller's ratchet is
a stochastic effect that occurs when either the population size
becomes too small, or the mutation rate too high, to sustain a finite
population in equilibrium. The population then starts to lose the
highest fitness individuals, a process that continues unabated until
the population dies out~\cite{LynchGabriel90}. In contrast to that,
here we do not observe a continued loss of fitness, we rather find
that the population sustains itself in the neighborhood of high local
optima. The difference between our situation and Muller's ratchet
model is that the ratchet model assumes a single peak landscape, where
only extremely rare back mutations can compensate for the loss in
fitness that the forward mutations carry. In a rugged landscape such
as the NK landscape, on the other hand, there are a large number of
local optima nearby, which leads to a significant number of
compensatory mutations (which, in effect, delocalize the
population). Therefore, although the population becomes delocalized,
it can nevertheless evade continued fitness loss and extinction.

Let us now discuss the implications of our findings for evolutionary
optimization. As we have mentioned above, in evolutionary optimization
we are interested in the highest possible maximum fitness in the
population. We have found that in NK landscapes, this is realized for
mutation rates close to the delocalization transition. This could be
utilized for evolutionary optimization in the following way. A number
of short initial runs could be used to find the regime in which the
delocalization transition takes place. Production runs would then be
run at a mutation rate slightly below that regime.

We should also point out that the genomic mutation rates found for the
NK landscape are by no means meant to be universal. We expect to find
different rates for different fitness landscapes, although we expect
the order of events (highest mean fitness, highest maximum fitness,
delocalization, error threshold) with increasing mutation rate to
remain the same (some of the events may coincide). For example, it is
suspected that the mutation rate of RNA viruses is optimal at around
$\mu\approx0.76$~\cite{Drake99} while it appears to be much lower for
DNA viruses ($\mu\approx 0.003$~\cite{drake91}). The optimal rate for
finding fitness maxima in the ``Royal Road''
landscape~\cite{vanNimwegen00} was found to be of the order of
$\mu=0.5$, and around $\mu=1.0$ for digital organisms~\cite{Adami98}.
In general, optimal rates depend on the supply of advantageous
mutations, and on the average height of maxima with respect to random
sequences. The presence of a delocalization transition depends on
whether or not a fair supply of ``secondary'' maxima can be found near
the initial fitness peak.

It is interesting to compare the delocalization transition of the
present work to the behavior of an adaptive walker in a dynamic
fitness landscape. In Ref.~\cite{WilkeMartinetz99}, it was shown that
an adaptive walker can traverse the entire sequence space if the
fitness landscape is changing slowly.  The average fitness encountered
on such walks (from peak to peak) lies above the average height of
local optima in such landscapes if the landscape changes slowly
enough.  Here, we find that a population can even traverse a static
landscape if the mutation rate is sufficiently high.  The local maxima
visited by the population (after delocalization) are of similarly high
fitness, as can be seen from Fig.~\ref{fig:meanmax-and-hamming}a) for
$u=0.01$.  The main difference between the two processes is that the
adaptive walker corresponds to a population evolving at a very low
mutation rate, such that the mean fitness and the maximum fitness in
the population coincide. The high mutation rate necessary for
delocalization, on the other hand, implies that the average fitness
lies far below the maximum fitness. Therefore, it is unclear to what
extent a natural population could utilize these local optima, if at
the same time the population has to suffer from a high load of
deleterious mutations. Nevertheless, as we have mentioned above, these
local optima may present an effective means to halt Muller's ratchet.

\section*{Acknowledgments}

We would like to thank J. F. Fontanari for helpful discussions, and
also for pointing out reference \cite{BonhoefferStadler93}. This work
was initiated during a stay of PRAC at the California Institute of
Technology.  PRAC is supported by Funda\c{c}\~ao de Amparo \`a
Pesquisa do Estado de S\~ao Paulo ({\bf FAPESP}). COW and CA are
supported by the NSF under contract No. DEB-9981397. The work of CA
was carried out in part at the Jet Propulsion Laboratory, under a contract
with the National Aeronautics and Space Administration.

\begin{appendix}

\section{Auto-correlation function in NK landscapes}

The auto-correlation function $\rho(d)$ at Hamming distance $d$
can be calculated with the following reasoning: we only have to
calculate the probability that the fitness contribution of a single gene
remains unchanged after $d$ mutations~\cite{Fontanaetal93}. For $d=1$, this
probability is $(N-K)/N$, since there are $N-K$ of the total $N$ genes that
will not affect the state of the particular gene we are interested in. For
$d=2$, we have then $(N-K)(N-K-1)/[N(N-1)]$, since the second mutation may hit
any of the $N-K-1$ genes among the $N-1$ remaining ones. Clearly, for
arbitrary $d$, we have
\begin{eqnarray}\label{nk-correlation}
  \rho(d) &=& \frac{(N-K)(N-K-1)\dots(N-K-d+1)}{N(N-1)\dots(N-d+1)}\nonumber\\
&=& \frac{(N-K)!(N-d)!}{N!(N-K-d)!}\,.
\end{eqnarray}
Note that this derivation does not depend on whether the $K-1$ interacting
genes are chosen randomly (as we assumed throughout the present paper) or as
the nearest neighbors of the gene they interact with. Even if all $K$ genes
are chosen completely randomly for every $f_i$ (the ``purely random'' version
of the NK model), the result remains unaltered. 

Equation (\ref{nk-correlation}) differs from previously reported results for
the autocorrelation function in NK landscapes.  In~\cite{Fontanaetal93}, three
different functional forms for $\rho(d)$ are reported for the three different
types of NK landscapes.  Similar results are given
in~\cite{Weinberger91,WeinbergerStadler93,SchusterStadler94}. The results
given in~\cite{Fontanaetal93} for the random neighbor and purely random model
are of a very simple functional form, and clearly disagree with
Eq.~(\ref{nk-correlation}). The result for the nearest neighbor model, on the
other hand, is more complicated and involves a sum over combinatorial terms. A
detailed analysis reveals that the sum can actually be taken, and the
resulting expression simplifies to ours. Hence, the previously published
correlation function for the nearest neighbor case is correct, though awkward,
while the other two cases are incorrect.

\end{appendix}
\bibliographystyle{unsrt}
\bibliography{paper.bib}

\end{document}